\documentstyle[12pt,psfig]{article}
\def\be{\begin{equation}}
\def\ee{\end{equation}}
\def\bea{\begin{eqnarray}}
\def\eea{\end{eqnarray}}
\def\lsim{\:\raisebox{-0.5ex}{$\stackrel{\textstyle<}{\sim}$}\:}
 
\begin{document}
\begin{flushright}
TIFR/TH/01-07 \\
SINP/TNP/01-07\\
\end{flushright}
\bigskip
\begin{center}
{\bf Reviving the energy independent suppression of the solar neutrino
flux} \\[2cm]
Sandhya Choubey$^a$, Srubabati Goswami$^a$, Nayantara Gupta$^b$,
D.P. Roy$^c$ \\[1cm]
$^a$ Saha Institute of Nuclear Physics, \\
1/AF, Bidhannagar, Calcutta 700 064, INDIA \\
$^b$ Indian Association for the Cultivation of Sciences \\
Calcutta 700 032, INDIA \\
$^c$ Tata Institute of Fundamental Research \\
Homi Bhabha Road, Mumbai 400 005, INDIA
\end{center}

\vspace{1cm}

\begin{center}
{\bf Abstract}
\end{center}
\bigskip

We explore the possibility of an energy independent suppression of the
solar neutrino flux in the context of the recent SuperKamiokande data.
From a global analysis of the rate and spectrum data, this scenario is
allowed at only 14\% probability with the observed Cl rate.  If we
allow for a 20\% upward renormalisation of the Cl rate along with a
downward renormalisation of the $B$ neutrino flux then the fit
improves considerably to a probability of $\sim 50\%$.  We compare the
quality of these fits with those of the MSW solutions.  These
renormalisations are also found to improve the quality of the fits
with MSW solutions and enlarge the allowed region of their validity in
the parameter space substantially.  Over much of this enlarged region
the matter effects on the suppression of the solar neutrino flux are
found to be very weak, so that the solutions become practically energy
independent.
\bigskip\bigskip

The results from the SuperKamiokande (SK) continue to confirm the
suppression of the solar neutrino flux as compared to the standard
solar model prediction [1].  The result from the GNO experiment [2]
is consistent with the earlier Ga experiment results from Gallex and
SAGE [3].  The most popular explanation to this suppression is
neutrino oscillation either in vacuum or in matter.  Table 1 shows the
suppression rate or survival probability of the solar neutrino
$(P_{\nu_e \nu_e})$ from the combined Ga [2,3], Cl [4] and SK[1] experiments
along with their energy thresholds.  The corresponding compositions of
the solar neutrino flux are also indicated.  The SK suppression rate
shown in brackets is appropriate for the oscillation of $\nu_e$ into
another active neutrino $(\nu_{\mu,\tau})$.  It is obtained by
subtracting the neutral current contribution of $\nu_{\mu,\tau}$ from
the SK rate.
\bigskip
\[
\begin{tabular}{|c|c|c|c|}
\hline
Experiment & Gallium & Clorine & SuperKamiokanda \\
\hline
&&& \\
Suppr. Rate & 0.576 $\pm$ 0.04 & 0.327 $\pm$ 0.029 & 0.465 $\pm$ 0.015 \\
&&& (0.36 $\pm$ .015) \\
&&& \\
${\rm E}_{\rm th}$ (MeV) & 0.2 & 0.8 & 6.5 \\
&&& \\
Composition & $pp$ (55\%), $Be$ (25\%), $B$ (10\%) & $B$ (75\%), $Be$
(15\%) & $B$ (100\%) \\
&&& \\
\hline
\end{tabular}
\]
Table 1. The suppression rates of solar neutrino $P_{\nu_e
\nu_e}$ shown for Ga, Cl and SK experiments [1-6] along with their
threshold energies and compositions.  The effective suppression rate
of the SK experiment, appropriate for $\nu_e \rightarrow
\nu_{\mu,\tau}$ oscillation, is shown in bracket.  All the suppression
rates are shown relative to the standard solar model prediction of
BP00 [7].
\bigskip

The observed energy dependence in the suppression rates in Ga, Cl and
SK experiments can be explained by the vacuum oscillation (VO), small
and large mixing angle MSW (SMA and LMA) as well as the LOW solutions
[5].  The apparent energy dependence comes from assuming the sun-earth
distance to coincide with the oscillation node of a MeV range neutrino
in the VO solution, while it comes from matter effects in the sun for the MSW
solutions and in the earth for the LOW solution.  The VO and SMA
solutions show strong and nonmonotonic energy dependence.  But the LMA
and LOW solutions show monotonic decrease from Ga to SK energies in
contrast to the apparent rise between the Cl and SK rates.

The recent SK data on the energy spectra at day and night show no
evidence of any energy dependence nor any day-night asymmetry in the
suppression rate [1].  This rules out a large part of the parameter
space.  In particular it practically rules out the VO solution and
disfavours the SMA along with a part of the LMA solution [1,6].  The
remaining part of the LMA solution and the low solution show
relatively weak energy dependence and cover two small patches in the
parameter space. Thus it is possible to explain the above mentioned
rates and the energy spectra only for very limited ranges of neutrino
mass $\Delta m^2$ and mixing angle $\theta$.

We present here fits to the above data, first with an energy
independent solution and then with two-flavour oscillation including
matter effects.  We shall see that with reasonable allowance for the
renormalisations of the Cl rate and the $B$ neutrino flux the data are
described well by the energy independent solution.  These
renormalisations shall also be seen to improve the quality of fits with
the oscillation solutions and enlarge the region of their validity in
the parameter space substantially.  Moreover we shall see that most of
this enlarged region of parameter space shows weak matter effect on
$P_{\nu_e \nu_e}$, implying practically energy independent suppression
of the solar neutrino flux.  Thus the energy independent solution can
be looked upon as an effective parameterisation of the oscillation
solutions over this region.

The definition of $\chi^2$ used in our fits is,
\begin{equation}
\chi^2 = \sum_{i,j} \left(F_i^{th} -
F_i^{exp}\right)
(\sigma_{ij}^{-2}) \left(F_j^{th} - F_j^{exp}\right)
\label{two}
\end{equation}
Where i,j runs over the number of experimental data points.  Here
$F_{i}^{\alpha}= {T_i^{\alpha}}/{T_{i}^{BP00}}$ where $\alpha$ is the
theoretical prediction or the experimental value of the event rate,
normalised by the standard solar model prediction of BP00 [7].
$F_{i}^{exp}$ is taken from Table 1 for total rates and from [8] for
day/night spectrum.  The error matrix $\sigma_{ij}$ contains the
experimental errors, the theoretical errors and their correlations.
For evaluating the error matrix we use the procedure described in [9].
The details of the solar neutrino code used is described in [10,11].
As in [12] we vary the normalisation of the spectrum as a free
parameter which avoids the overcounting of the rates and spectrum data
for SK. Hence for the day/night spectrum analysis we have (36 - 1)
degrees of freedom (d.o.f) while for the total rates we have 3, which
makes a total of 38 d.o.f for the rates+spectrum analysis.  In
addition to the best-fit parameter values and $\chi^2_{min}$ we shall
present the goodness of fit (g.o.f.) of a solution where by g.o.f. we
mean the probability that the $\chi^2$ will exceed the $\chi^2_{min}$
for a correct model.  Finally for the general oscillation solution
with matter effects, we shall also delineate the 90\%, 95\% and 99\%
allowed regions in the two parameter $\Delta m^2 - \tan^2 \theta$
plane. These regions are defined as $\chi^2 \leq \chi^2_{min} + \Delta
\chi^2$ where $\Delta \chi^2$ is 4.61,5.99,9.21 respectively for two
parameters and the $\chi^2_{min}$ corresponds to the global $\chi^2$
minimum [13].
\bigskip

\noindent {\bf Energy Independent Solution:} 
\medskip

\nobreak
We shall explore first the possibility of explaining these data in
terms of a simple energy independent solution by assuming modest
changes in the Cl rate and the $B$ neutrino flux.  Energy independent
solutions to the solar neutrino anomaly have been considered by
several authors in the past [14].  Traditionally it is associated with
the vacuum oscillation solution at a distance much larger than the
oscillation wave-length, so that the average survival probability \be
P_{\nu_e \nu_e} = 1- {1 \over 2} \sin^2 2\theta.
\label{one}
\ee 
We shall see below however that this survival probability remains
approximately valid over a wide range of parameters even after
including the matter effects in the sun and the earth, since the
matter effects over this range are too small to be measurable at the
present level of experimental accuracy.  In particular the so called
bimaximal mixing solution, corresponding to nearly maximal mixing of
solar neutrino $(\nu_e)$, implies such an energy independent solution
over a very wide range of $\Delta m^2$ [15].  Moreover an energy
independent solution to the solar neutrino anomaly seems to offer the
possibility of explaining the atmospheric and the LSND neutrino
anomalies as well without assuming any sterile neutrino [16].

We present the results of our fit to the combined data on rates and
the SK day/night spectrum with the energy independent solution (2)
in Table 2 for both $\nu_e$ oscillations into active and sterile
neutrinos.  In order to reconcile the energy independence of the
spectrum with the apparent energy dependence in the rates of Table 1,
we have cosidered the following changes in the Cl rate and $B$
neutrino flux.

\begin{enumerate}

\item[{(A)}] Clorine Rate: Since the Cl experiment [4] has not been
calibrated, there are several fits in the literature disregarding this
rate [6,17,18].  In any case the apparent rise between the Cl and SK
rates is in direct conflict with the predicted fall for the LMA and
LOW solutions, which are favoured by the spectrum data.  Therefore we
have considered an upward renormalisation of the Cl data by 20\%,
which is a $2\sigma$ effect.

\item[{(B)}] Boron Neutrino Flux: The $B$ neutrino flux is very
sensitive to the solar core temperature and hence to the underlying
solar model.  Even within the standard solar model estimate of BP00 it
has a large error bar, i.e.
\be
f_B = 5.15 \times 10^6/{\rm cm}^2/\sec \left(1.0 \matrix{+.20 \cr
-.16}\right). 
\label{three}
\ee
Therefore we have considered a variation of this flux within a
corridor of about $\pm 2\sigma$ of the above central value.
\end{enumerate}
\vskip 5pt
\begin{table}[htb]
\begin{center}
\begin{tabular}{|c|c|c|c|c|c|}
\hline
&Nature of &$X_{B}$&$\sin^22\theta\left(\matrix{\tan^2\theta \cr {\rm
or} \cr \cot^2 \theta}\right)$& $\chi^2_{\min}$ & Goodness   \\
&Solution & & & &of fit \\
& & &  & &\\  
\hline 
& & 1.0&0.95(0.63)& 46.15 &14.39\%\\ \cline{3-6}
Chlorine&Active &0.77 & 0.96(0.68) & 45.86 & 12.25\%\\ \cline{2-6}
Observed&&1.0 &0.89(0.52) &53.66 &3.75\%\\
\cline{3-6} 
&Sterile  &0.83&0.92(0.57)&54.63& 2.55\%\\ 
\hline 
&&1.0 &0.89(0.52) &37.65& 43.13\%\\ \cline{3-6}
Chlorine&Active &0.73&0.88(0.50)&36.05 & 46.63\%\\ \cline{2-6}
Renormalised&&1.0&0.85(0.44)&40.89&30.35\%\\
\cline{3-6}
&Sterile&0.79&0.85(0.44)&40.09&26.4\%\\
\hline 
\end{tabular}
\end{center}
\begin{description}
\item Table 2: The best-fit value of the parameter, the $\chi^2_{\min}$
and the g.o.f from a combined analysis of rate and spectrum with the
energy independent solution (2).
\end{description}
\end{table}

Table 2 shows that taking the experimental rates at their face value
results in a poor fit to the energy independent solution,
corresponding to a probability of 14.39 (3.75)\% for the active
(sterile) case.  But assuming a 20\% renormalisation of the Cl rate and
floating the normalisation of $B$ flux improves the probability to 46.6
(26.4)\%.  Let us comment on two related features of this fit, which
may appear counter intuitive.  The theoretical survival rate from
(2) is $\geq 0.5$ while the experimental rates from Cl and SK with the
BP00 $B$ neutrino flux are significantly lower than 0.5 (Table 1).
Thus one would naively expect the fit with the BP00 neutrino flux,
denoted by $X_B = 1$, to result in $\sin^2 2\theta = 1$ and a much
larger $\chi^2_{min}$ than the free $X_B$ fit.  We have checked these
to be true if we drop the theoretical error in (1), reducing it to the
standard expression for $\chi^2$.  However including the large
uncertainty in the $B$ neutrino flux of (3) via the theoretical error matrix
implies that the best fit with the BP00 flux ($X_B = 1$ solution)
corresponds actually to an $X_B$ significantly lower than 1.  Hence the
corresponding $\chi^2_{min}$ and $\sin^2 2\theta$ values are close to
those of the free $X_B$ fit.  
The $\chi^2_{min}$ is found to be quite flat in $\sin^2 2\theta$ (and
even more so in $\tan^2 \theta$)  in the region
around the best-fit values.  It may be noted here that using 1$\sigma$
lower limits of the appropriate nuclear reaction rates Brun,
Truck-Chieze and Morel [19] have obtained a relatively low value of
$B$ neutrino flux, \be f_B = 3.21 \times 10^6/{\rm cm}^2/\sec,
\label{four}
\ee
and found it to give better agreement to the helioseismic data than
(3).  This corresponds to a low value of $X_B \simeq 0.63$, 
which is about $2\sigma$ below the central value of (3).  More
recently a $X_B \simeq 0.75$ has been 
obtained from the helioseismic model using standard values of the
nuclear reaction rates [20].
\bigskip

\noindent {\bf Oscillation Solution with Matter Effects:}
\medskip

\nobreak
We shall first analyse the two-flavour oscillation solution, including
the matter effects in the sun and the earth, to determine the region of
the $\Delta m^2 - \tan^2\theta$ plane in which it effectively reduces
to the energy independent solution (2).  Since the normalisation
uncertainty of Ga and Cl experiments are $> 10\%$ each, one cannot
experimentally distinguish the solutions showing energy dependence of
$< 10\%$ over the energy range of Ga to SK experimeriments from the
energy independent solution (2).  We have therefore assumed that a
matter effect of $< 10\%$ on $P_{\nu_e\nu_e}$ over the range of Ga to
SK neutrino energies is a good working  
definition for an effectively energy independent solution (2).  Fig. 1
shows the region in the $\Delta m^2 - \tan^2\theta$ plane, where the
oscillation solution including matter effects effectively reduces to
the energy independent solution (2).  We have restricted the plot to
$\Delta m^2 < 10^{-3} \ {\rm eV}^2$ in view of the severe constraint
from the CHOOZ experiment above this range [21].  One sees from Fig. 1
two distinct regions of validity of the energy independent solution
(2). Firstly the solar matter effect is negligible in the shaded
region above the MSW range [17] 
\be
10^{-15} \ {\rm eV} \lsim \Delta m^2/4E \lsim 10^{-11} \ {\rm eV},
\label{five}
\ee
with $4E \sim 1 - 50$ MeV, along with the triangular region below.
Moreover a narrow strip around $\tan^2
\theta = 1$ represents the region of near maximal-mixing, where the
MSW solution for $P_{\nu_e\nu_e}$ reduces to the vacuum solution (2).
However the regeneration effect in earth makes a significant
contribution to the maximal-mixing region over $\Delta m^2 \simeq 10^{-5} -
10^{-7} \ {\rm eV}^2$, which accounts for the gap in this strip. It
is this near maximal-mixing strip that is relevant for the energy independent
solution of Table 2 and the oscillation solutions presented below.

We shall now present the fits of the oscillation solutions including matter
effects to the combined data on rates and the SK day/night spectrum.
As in the previous case we have done these fits with the rates shown
in Table 1 as well as those with renormalised Cl rate and $B$ neutrino
flux. 

\begin{center}
\begin{tabular}{|c|c|c|c|c|c|c|}
\hline
&Type of & Nature of & $\Delta m^2$ &
$tan^2\theta$ & $\chi^2_{min}$& g.o.f\\
&Neutrino & Solution & in eV$^2$& & &\\
\hline
& & SMA & $5.48 \times 10^{-6}$&$5.79\times 10^{-4}$ 
&43.22 &19.01\%  \\ \cline{3-7}
&Active & LMA & $4.18\times 10^{-5}$ &0.36&
37.33 & 40.78\%  \\ \cline{3-7} 
Cl& & LOW & $1.51\times 10^{-7}$ & 0.64 &39.54&31.48\%  \\ \cline{2-7} 
Obsvd.& & SMA & $3.74 \times 10^{-6}$&$5.2 \times 10^{-4}$ &
 44.85 & 14.79\%  \\ \cline{3-7}
&Sterile & LMA & $1.03 \times 10^{-4}$ & 0.58 &
52.18 & 3.96\%  \\ \cline{3-7} 
& & LOW & $3.47 \times 10^{-8}$ & 0.82 & 50.57&5.43\%  \\ \hline 

& & SMA & $4.97 \times 10^{-6}$& $3.15 \times 10^{-4}$
& 42.43& 21.37\%  \\ \cline{3-7}
Cl&Active & LMA & $6.68\times 10^{-5}$ &0.39&30.32& 73.53\%\\ 
\cline{3-7} 
Renorm.& & LOW & $1.63 \times 10^{-7}$ &0.76& 32.43& 63.9\% \\ \cline{2-7} 
& & SMA & $3.44 \times 10^{-6}$& $3.59 \times 10^{-4}$
 &41.98&22.76\%  \\ \cline{3-7}
&Sterile & LMA & $1.04 \times 10^{-4}$ &0.53&37.9& 38.27\%\\ 
\cline{3-7} 
& & LOW & $4.08 \times 10^{-8}$ & 0.84 & 36.8 & 42.98\%\\ \hline 
\end{tabular}
\end{center}
\begin{description}
\item Table 3: The best-fit values of the parameters, the $\chi^2_{\min}$
and the g.o.f from a combined analysis of rate and spectrum in terms
of $\nu_e$ oscillation into an active/sterile neutrino, including the
matter effects.
\end{description}

Table 3 summarises the results of fitting the oscillation solutions to
the combined data with the observed and renormalised Cl rates assuming
the $B$ neutrino flux of BP00 $(X_B = 1)$.  We see from the upper part
of this table that one can get acceptable fits to the data taken at
its face value in terms of $\nu_e$ oscillation into active flavour in
the LMA and LOW regions, while the SMA region gives a marginal fit at
19\% probability.  The corresponding oscillation solutions into
sterile neutrino give poor fits.  Renormalising the Cl rate upwards by
20\% improves the quality of oscillation solutions into active flavour
remarkably in the LMA and LOW regions but not in the SMA region. The
quality of the sterile solutions also improves to acceptable levels of
probability; but they remain inferior to the oscillation solutions
into active neutrino.  

A clear pedagogical discussion of the LMA and LOW solutions can be
found in [17].  We shall only point out here some essential features,
focussing on the oscillation into active neutrino.  The $\Delta m^2$
of the LMA solutions of Table 3 lie at the upper edge (adiabatic edge)
of the MSW range (5).  They lie outside this range for Ga energy, so
that the corresponding survival rate is approximated by (2), or equivalently 
\be
P_{\nu_e\nu_e} \approx {1 \over 2} (1 + \epsilon^2),
\label{six}
\ee
\be
\epsilon = \cos 2\theta = {1 - \tan^2 \theta \over 1 + \tan^2 \theta}.
\label{seven}
\ee
On the other hand they lie inside the MSW range at SK energy.  Here
the solar $\nu_e$ gets adiabatically converted into the heavier one of
the two neutrino mass states,
\be
\nu_2 = \sin \theta \cdot \nu_e + \cos \theta \cdot \nu_{\mu,\tau}.
\label{eight}
\ee
The resulting survival probability on earth is
\be
P_{\nu_e \nu_e} \approx \sin^2 \theta = {1\over 2} (1 - \epsilon).
\label{nine}
\ee
With the LMA mixing angles of Table 3, corresponding to $\epsilon
\simeq 0.4$, eqs. (6) and (9) can be seen to roughly reproduce the Ga and SK
rates of Table 1.  In the LOW solutions the energy dependence arises
from the $\nu_e$ regeneration in earth during the night.  This is
known to be small for the SK energy from the absence of day/night
asymmetry, so that the corresponding rate (9) is valid for the LOW
solution as well.  However the regeneration contribution can be
significant for the LOW solution at Ga energy, which has moved into
the MSW range (5).  The corresponding rate after averaging over day
and night is [17]
\be
\bar{P}_{\nu_e \nu_e} = {1 \over 2} (1 - \epsilon + f_{reg}).
\label{ten}
\ee
For a constant density earth
\be
f_{reg} = {\eta_E (1 - \epsilon^2) \over 2(1 - 2\epsilon \eta_E +
\eta^2_E)}.
\label{eleven}
\ee
Thus the earth regeneration contribution is always positive and can be
significant for $\eta_E \sim 1$, where
\be
\eta_E = 0.66 \left({\Delta m^2/E \over 10^{-13} \ {\rm eV}}\right)
\left({g/{\rm cm}^3 \over \rho Y_e}\right).
\label{twelve}
\ee
Here $\rho$ is the matter density in earth and $Y_e$ the average
number of electrons per nucleon.  The maximal contribution comes from
$\Delta m^2 \sim 3E \times 10^{-13} \ {\rm eV} \sim 10^{-7} \ {\rm
eV}^2$ for Ga energy.  The mixing angle of the LOW solutions,
$\epsilon \sim 0.2$, corresponds to a maximal $f_{reg} \sim 0.3$,
which can account for the Ga rate of Table 1.
While
this description of the matter effects in the sun and the earth is
admittedly simplistic we have treated them rigorously in our calculation
at all energies.

Fig. 2 shows the 90\%, 95\% and 99\% C.L. allowed regions in the
$\Delta m^2 - \tan^2 \theta$ plane for the oscillation solutions into
active neutrino.  
We find that SMA solution is disallowed at 95\% (99\%)
C.L. with the observed (renormalised) Cl rate.  The allowed regions of
the LMA and LOW solutions increase mildly with the upward renormalisation of
the Cl rate.  Comparison of Figs. 1 and 2 shows that modest parts
of the allowed regions for both the LMA and LOW solutions correspond
to the effectively energy independent solution (2).

Table 4 summarises the effects of changing the $B$ neutrino flux
$(X_B)$ on the oscillations solutions into active neutrino.  It lists
the best solutions for $X_B = 0.75$, favoured by helioseismic model
[20], as well as for free $X_B$.  In each case the solutions are shown
for both observed and renormalized Cl rates.  It may be noted that the
$X_B = 0.75$ lies within $\sim 1.5 \sigma$ of the BP00 flux (3).  In
combination with the 20\% renormalisation of the Cl rate it would
imply that all the suppression rates of Table 1 agree with one another
within $1.5\sigma$.  Therefore this combination is expected to favour
an effectively energy independent solution.

\begin{center}
\begin{tabular}{|c|c|c|c|c|c|c|}
\hline
& & Nature of & $\Delta m^2$ &
$tan^2\theta$ & $\chi^2_{min}$& g.o.f\\
& & Solution & in eV$^2$&  & &\\
\hline
& & SMA & $5.43 \times 10^{-6}$&$5.09\times 10^{-4}$ 
&39.50 &31.64\%  \\ \cline{3-7}
&$X_B=0.75$ & LMA & $4.39\times 10^{-5}$ &0.54&
43.18 & 19.13\%  \\ \cline{3-7} 
Cl& & LOW & $1.41\times 10^{-7}$ & 0.69 &41.88&23.08\%  \\ \cline{2-7} 
Obsvd.&~~~~~0.57 & SMA & $5.35 \times 10^{-6}$&$4.35\times 10^{-4}$ &
 37.98 & 37.92\%  \\ \cline{3-7}
&$X_B \ 1.34$ & LMA & $4.21 \times 10^{-5}$ & 0.25 &
34.22 &55.34\%  \\ \cline{3-7} 
&~~~~~0.93 & LOW & $1.51 \times 10^{-7}$ & 0.63 & 39.59&31.28\%  \\ \hline 
& & SMA & $4.95 \times 10^{-6}$& $3.11 \times 10^{-4}$
& 38.15& 37.19\%  \\ \cline{3-7}
Cl&$X_B=0.75$ & LMA & $6.92\times 10^{-5}$ &0.57&34.06& 56.11\%\\ 
\cline{3-7} 
Renorm.& & LOW & $1.59 \times 10^{-7}$ &0.82& 32.76& 62.35\% \\ \cline{2-7} 
&~~~~~0.53 & SMA & $4.90 \times 10^{-6}$& $2.92 \times 10^{-4}$ &35.57&48.89\%  
\\ \cline{3-7}
&$X_B \ 1.14$ & LMA & $6.57 \times 10^{-5}$
&0.35&29.94& 75.14\%\\  
\cline{3-7} 
&~~~~~0.88 & LOW & $1.64 \times 10^{-7}$ & 0.76 & 31.95& 66.17\%\\ \hline 
\end{tabular}
\begin{description}
\item Table 4: Best fits to the combined rates and spectrum data in
terms of $\nu_e$ oscillation into active neutrino with $X_B = 0.75$
and free $X_B$.
\end{description}
\end{center}

As we see from the top part of Table 4, with observed Cl rate and $X_B
= 0.75$ the SMA gives a better fit than the LMA and LOW solutions.
This is because reducing $X_B$ to 0.75 accentuates the rise between
the Cl and the SK rates of Table 1 as it enhances the latter by a
larger amount.  Hence it favours the nonmonotonic energy dependence of
SMA over the monotonically decreasing energy dependence of LMA and LOW
solutions.  However this anomaly disappears with the upward
renormalisation of the Cl rate, so that the LMA and LOW solutions
become much better than the SMA.  Note also that reducing $X_B$ to
0.75 results in reducing the energy dependence between the Ga and SK
rates, resulting in larger mixing angle for the LMA solutions than in
Table 3. 

The free $X_B$ fits yield $X_B = 0.5 - 0.6$ for the SMA and $X_B > 1$
for the LMA solutions.  This is because $X_B = 0.5 - 0.6$ enhances the
SK rate more than Cl as favoured by SMA, while $X_B > 1$ suppresses
the SK rate more than Cl as favoured by the LMA solutions.  Note
however that $X_B > 1$ magnifies the decrease between the Ga and SK
rates, resulting in a smaller mixing angle for the LMA solutions than
in Table 3.  But the shift is small for the renormalised Cl rate.  The
LOW solution gives the worst fit for the observed Cl rate, since its
weak energy dependene favours $X_B < 1$ which accentuates the rise
between the Cl and SK rates.  But with renormalised Cl the LMA and LOW
solutions become much better than the SMA.

Fig. 3 shows the 90\%, 95\% and 99\% C.L. allowed regions in the
$\Delta m^2 - \tan^2 \theta$ plane for the free $X_B$ fits.  With the
observed Cl rate the LMA region covers relatively small mixing angles
while the LOW region is marginal.  Consequently there is little
overlap with the energy independent region of Fig. 1.  However for the
renormalised Cl rate the LMA region expands to larger mixing
angles and masses.  The LOW region also covers a large range of mass.
Consequently there is a significant overlap with the energy
independent region of Fig. 1.  

Finally Fig. 4 shows the corresponding allowed regions for $X_B =
0.75$ fits with observed and renormalised Cl rates.  In the former
case there is a large allowed region at 90\% C.L. for the SMA
solution.  However in the latter case the 90\% C.L. region disappears
from the SMA solution, while covering a very large range of masses and
mixing angles for the LMA and LOW solutions.  Comparing this contour
with Fig. 1 shows that the bulk of the expanded 90\% C.L. region in
this case corresponds effectively to the energy independent suppresion
rate (2), as anticipated earlier.  Note in particular the expansion of
the 90\% CL allowed region of the LMA solution well into the so called
dark region, corresponding to $\tan^2\theta > 1$ $(\epsilon < 0)$.  In
this region the MSW prediction of energy dependence via eqs. (6,9)
changes its direction, which goes against the direction of the data.
Nonetheless the energy dependence becomes so mild with $X_B = 0.75$
that the 90\% CL region extends upto $\tan^2 \theta > 2$, which is
beyond the range shown in Fig. 1.  In other words the region of
effectively energy independent solution shown in Fig. 1 is a
conservative one.  The expansion of the 90\% CL region of the LOW
solution also shows remarkable overlap with the lower energy
independent strip of Fig. 1.  It may be added here that both these
strips go down to $\Delta m^2 \sim 10^{-9} \ {\rm eV}^2$, below which
one gets significant energy dependence from vacuum oscillation.

Let us conclude by briefly discussing whether some of the forthcoming
neutrino experiments will be able to discreminate between the energy
independent and the MSW solutions. In particular the SNO experiment[22]
is expected to provide both the charged current and neutral current scattering
rates over roughly the same energy range as SK. Thus the B neutrino flux
can be factored out from their ratio, CC/NC. For oscillation into active
neutrino corresponding double ratio 
$R_{cc}/R_{nc}$
is predicted to be larger than 0.5 for the energy
independent solution (eq.2) and smaller than 0.5 for the LMA and LOW
solutions (eq.9). We have calculated the best fit values of this ratio
for the energy independent solution of Table 2  and the LMA and LOW solutions
of Table 3 with renormalised Cl rate. The predicted ratios are shown in 
Table 5.With a sample of 5000 CC and 2000 NC events 
the total $1\sigma$ error for
this ratio at SNO is expected to be about 4\%[17]. 
This will be able to distinguish
the energy idependent solution clearly from the LMA and to a lesser extent
from the LOW solution. On the other hand the LOW solution predicts a large
Day-Night asymmetry of $>$ 10\% for the Be neutrino[17,18] at the Borexino[23]
and the KamLAND[24] experiments. This will be able to distingusish the LOW
from the LMA and the energy independent solutions. 
Lastly it should be noted
that the reactor neutrino data at KamLAND is expected to show oscillatory
behavior for the LMA solution[25], 
which will help to distinguish it from the LOW or a generic energy 
independent solution. 

\[
\begin{tabular}{|c|c|c|c|}
\hline
Nature of Solution & $\Delta m^2$ & $tan^2 \theta$ & $R_{cc}/R_{nc}$
\\
\hline
LMA &  $6.68 \times 10^{-5}$ & 0.39  & 0.32  \\
\hline
LOW & $1.63 \times 10^{-7}$ & 0.76 & 0.45 \\ \hline
energy independent & -& 0.52 & 0.55 \\ \hline 
\end{tabular}
\]
Table 5. The $R_{cc}/R_{nc}$ at SNO at the best-fit values 
for the LMA, LOW and energy independent solution for the renormalised 
Chlorine and $X_{B}$ fixed cases of Table 2 and Table 3.   

\bigskip

\noindent {\bf Summary:} 
\medskip

\nobreak
In summary the recent SK data on day/night spectrum is in potential
conflict with the apparent energy dependence in the suppression rates
observed in Ga, Cl and SK experiments. Including matter effects one can get acceptable
oscillation solutions to both rates and spectrum data only
over limited regions of mass and mixing parameters.  However
an upward renormalisation of the Cl rate by 20\% $(2\sigma)$ results
in substantial improvement of the quality of fit.  Moreover a downward
renormalisation of the $B$ neutrino flux by 25\% $(1.5\sigma)$ as
suggested by the helioseismic model enlarges the allowed region of the
parameter space substantially.  Over most of this enlarged region the
energy dependence resulting from the matter effects is too weak to be
discernible at the present level of experimental accuracy. Hence with
these renormalisations of the Cl rate and the $B$ neutrino flux the
data can be described very well by an energy independent solution.
\vskip 5pt
{\it Note added}: After this work was completed
the 1258 days SK data has appeared on the net[26]. We have checked that 
the results of our analysis do not show any significant changes with
the new data.

We thank Profs. H.M. Antia, G. Bhattacharya, S.M. Chitre,
K. Kar and A. Raychaudhuri  
for
discussions. 

\bigskip\bigskip

\begin{center}
{\bf References}
\end{center}
\bigskip

\begin{enumerate}
\item[{[1.]}] Y. Suzuki (SK collaboration), in Neutrino 200, Sudbury,
Canada (2000); The results of all the neutrino experiments presented
at the Neutrino2000 meeting can be found at http://nu2000.sno.laurentian.ca.
\item[{[2.]}] M. Altmann {\it et al.}, {\em Phys. Lett.} {\bf B490}, 16 (2000). 
\item[{[3.]}] W. Hampel {\em et al.}, (The Gallex collaboration), {\em
Phys. Lett.} {\bf B388}, 384 (1996);  J.N. Abdurashitov {\em et
al.}, (The SAGE collaboration), {\em Phys. Rev. Lett.} {\bf 77},
4708 (1996).
\item[{[4.]}] B.T. Cleveland et al., Astrophys. J. 496, 505 (1998);
R. Davis, Prog. Part. Nucl. Phys. 32, 13 (1994).
\item[{[5.]}] J.N. Bahcall, P.I. Kraslev and A.Y. Smirnov,
Phys. Rev. D58, 096016 (1998).
\item[{[6.]}] M.C.Gonzalez-Garcia, C. Pe\~{n}a-Garay,
Nucl.Phys.Proc.Suppl.{\bf 91},80 (2000).
\item[{[7.]}] J. N. Bahcall, M. H. Pinsonneault and Sarbani Basu, astro-ph/0010346.
\item[{[8.]}] The day/night spectrum data used is obtained from
Y. Suzuki of SK collaboration, private communication.
\item[{[9.]}] G.L. Fogli and E. Lisi, Astropart. Phys. {\bf 3},
185 (1995).
\item[{[10.]}] S. Goswami, D. Majumdar, A. Raychaudhuri, hep-ph/9909453;
Phys. Rev. {\bf D63},
013003, 2001.
\item[{[11.]}] A. Bandyopadhyay, S. Choubey, S. Goswami, hep-ph/0101273,
To appear in Phys. Rev. D.
\item[{[12.]}] M.C. Gonzalez-Garcia, P.C.  de Holanda, C.
Pe\~{n}a-Garay, and J.W.F. Valle, hep-ph/9906469, {\em Nucl. Phys.}
{\bf B573}, 3 (2000).
\item[{[13.]}] See e.g. Numerical Recipe in Fortran 77, W.H. Press et
al., Cambridge University Press (2nd edition, 1991).
\item[{[14.]}] 
A. Acker, S. Pakvasa, J. Learned and T.J. Weiler, Phys. Lett.
B298,149(1993);
P.F. Harrison, D.H. Perkins and W.G. Scott, Phys. Lett. B349,137(1995)
and B374,111(1996); R. Foot and R.R. Volkas, hep-ph/9510312;
A. Acker and S. Pakvasa, Phys. Lett. B397,209(1997);
P.I. Krastev and S.T. Petkov, Phys. Lett. B395,69(1997);
; G. Conforto et al., Phys. Lett. B427, 314 (1998); 
W.G. Scott, Nucl. Phys. B (Proc. Suppl.) 66, 411 (1998) and hep-ph/0010335.
\item[{[15.]}] Y. Grossman and Y. Nir, Nucl. Phys. B448, 30 (1995);
R. Barbieri, L. Hall, D. Smith, A. Strumia and N. Weiner, JHEP 9812,
017 (1998).
\item[{[16.]}] G. Barenboim and F. Scheck, Phys. Lett. 440B, 332
(1998).
\item[{[17.]}] M.C. Gonzalez-Garcia, C. Pena-Garay, Y. Nir and
A.Yu. Smirnov, {\em Phys.Rev.} {\bf D63}, 013007 (2001). 
\item[{[18.]}] Andre de Gouvea, A. Friedland, and H. Murayama,
{\em Phys.Lett.} , {\bf B490}, 125 (2000) 
\item[{[19.]}] A.S. Brun, S. Truck-Chize and P. Morel,
Astrophys. J. 506, 913 (1998).
\item[{[20.]}] H.M. Antia and S.M. Chitre, Astronomy and Astrophysics,
347, 1000 (1999); S. Goswami, K. Kar, H.M. Antia and S.M. Chitre, To
be published in Proc. Helio and Asteroseismology at the dawn of the
Millennium, Tenerife, Spain, 2-6 October, 2000.
\item[{[21.]}] M. Apollonio et al., Phys. Lett. B446, 415 (1999).
\item[{[22.]}]A.B. McDonald for the SNO collaboration, Nucl.Phys. Proc.
Suppl. {\bf 91}, 21, 2000:
\item[{[23.]}]G. Ranucii {\it et al.}, Borexino Collaboration, Nucl. Phys.
Proc.Suppl. {\bf 91}, 58, 2001.  
\item[{[24.]}]J. Busenitz {\it et al.}, "Proposal for US participation
in KamLAND", March 1999 (unpublished). 
\item[{[25.]}]A. Piepke for the KamLAND collaboration, Nucl. Phys. Proc. 
Suppl. {\bf 91}, 99, 2001; 
V. Barger, D. Marfatia, Wood.B.P, Phys. Lett. {\bf B498}, 53,
2001. 
\item[{[26.]}]S. Fukuda {\it et al.}, hep-ex/0103033.
\end{enumerate}

\centerline{\psfig{figure=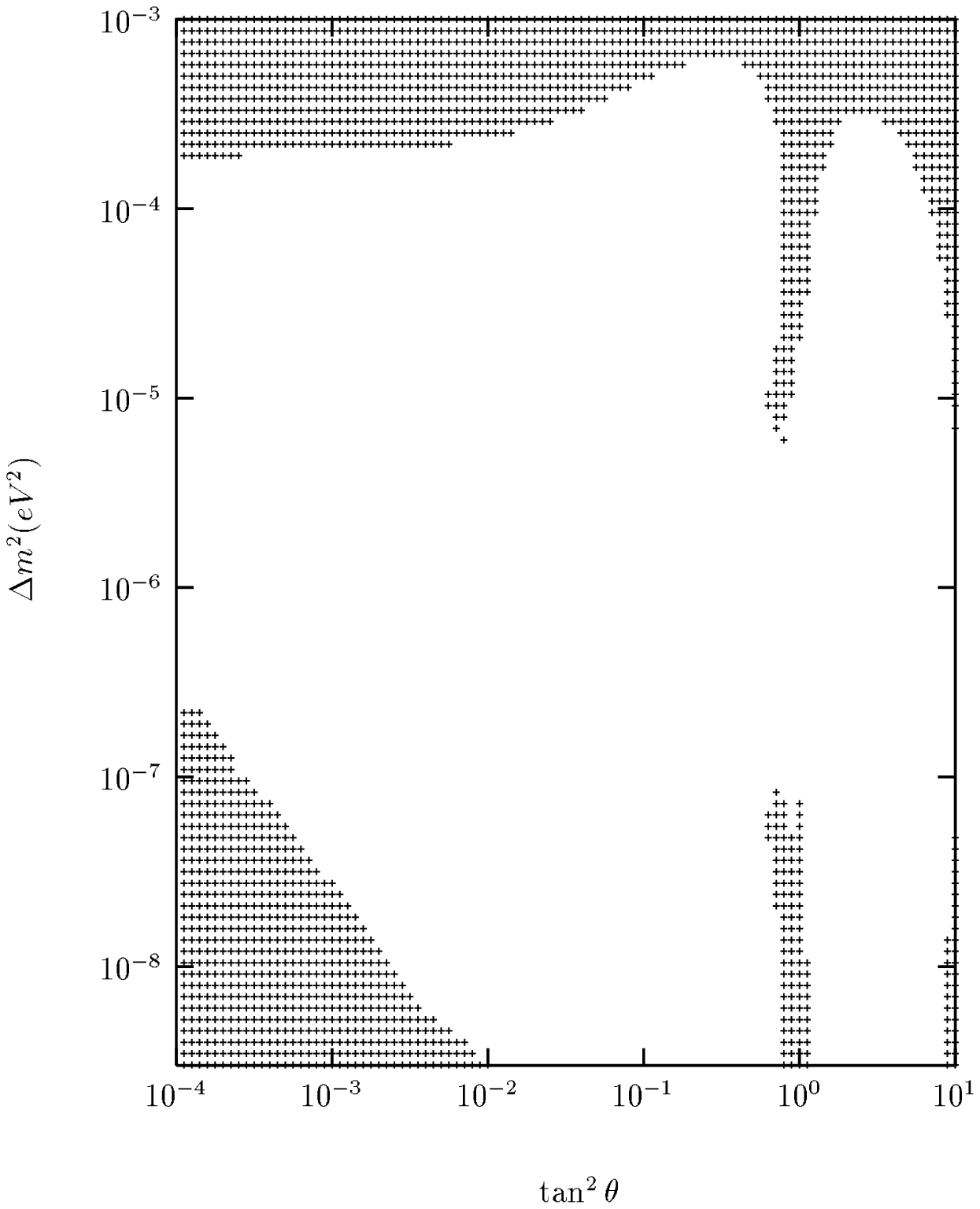,width=16.5cm,height=20.5cm}}
\vskip -1in
\parbox{5in}{
{\bf Fig 1} The quasi energy independent
allowed region in $\Delta m^2-\tan^2 \theta$ parameter
space where the solar neutrino survival probability agrees with 
eq. (2) to within 10\% over the Ga and SK energies.}  

\newpage
\topmargin -2.5in
\centerline{\psfig{figure=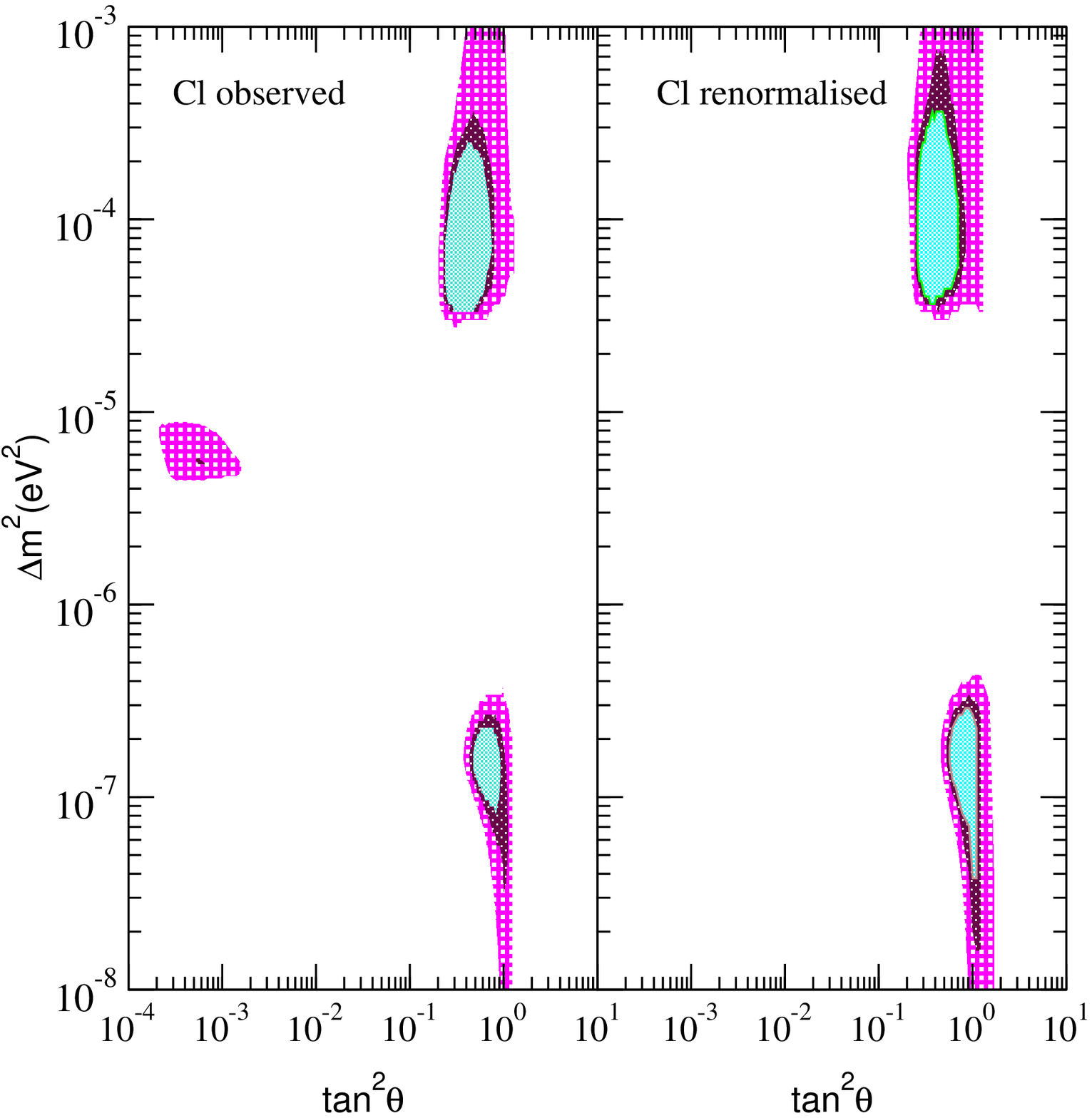,width=18.0cm,height=18.5cm}}
\vskip 0in 
\parbox{5in}{
{\bf Fig. 2}: The 90, 95 and 99\% C.L. allowed area from the
global analysis of the total rates from Cl (observed  and
20\% renormalised), Ga and SK detectors
and the 1117 days SK recoil electron spectrum at day and night,
assuming MSW conversions to active neutrinos.}

\newpage
\topmargin -1in
\centerline{\psfig{figure=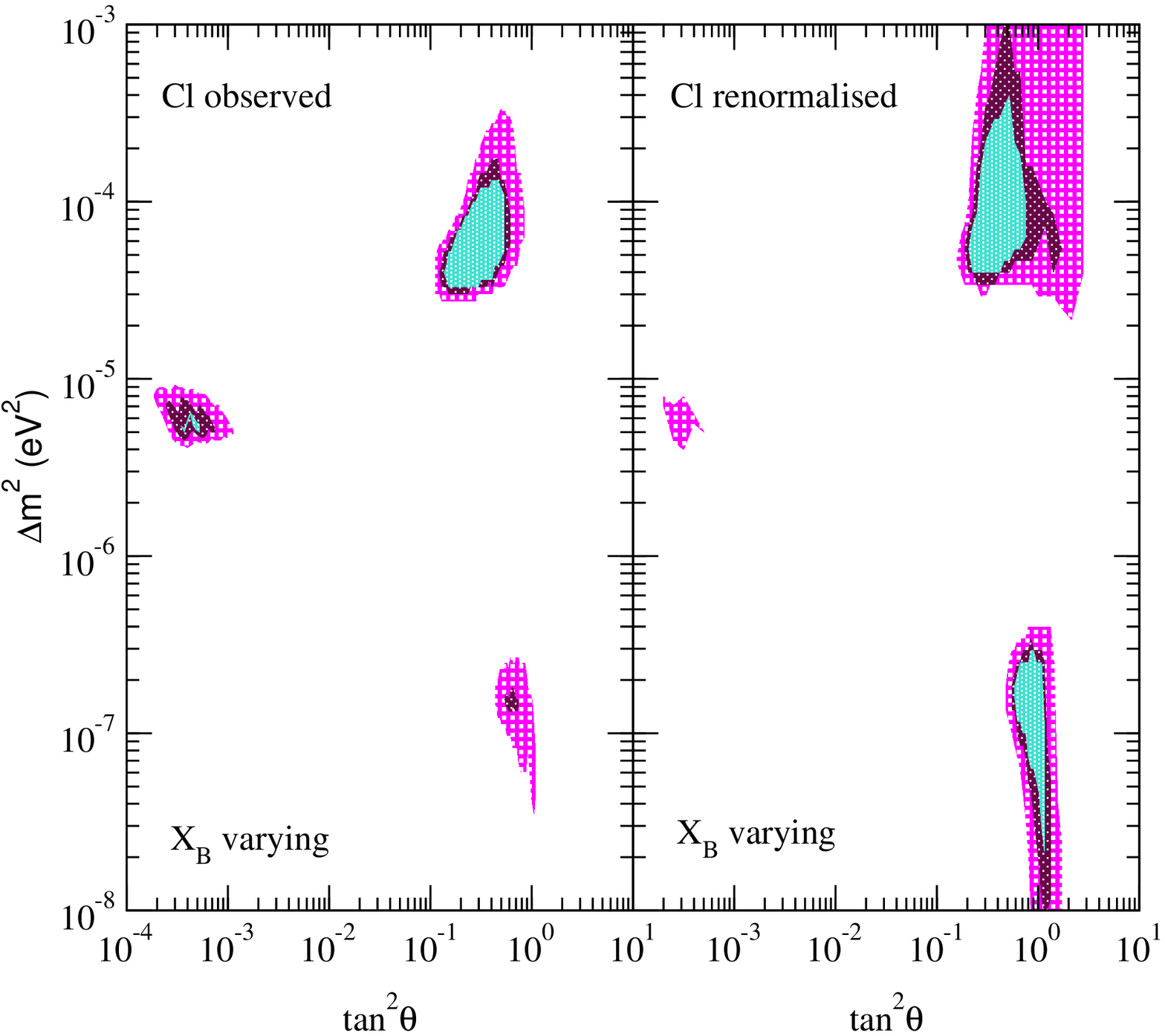,width=16.5cm,height=20.5cm}}
\vskip -1in
\parbox{5in}{
{\bf Fig. 3}: The 90, 95 and 99\% C.L. allowed area from the
global analysis of the total rates from Cl (observed  and
20\% renormalised), Ga and SK detectors
and the 1117 days SK recoil electron spectrum at day and night,
assuming MSW conversions to active neutrinos. The B normalisation
is floated as a free parameter.}

\newpage
\topmargin -1in
\centerline{\psfig{figure=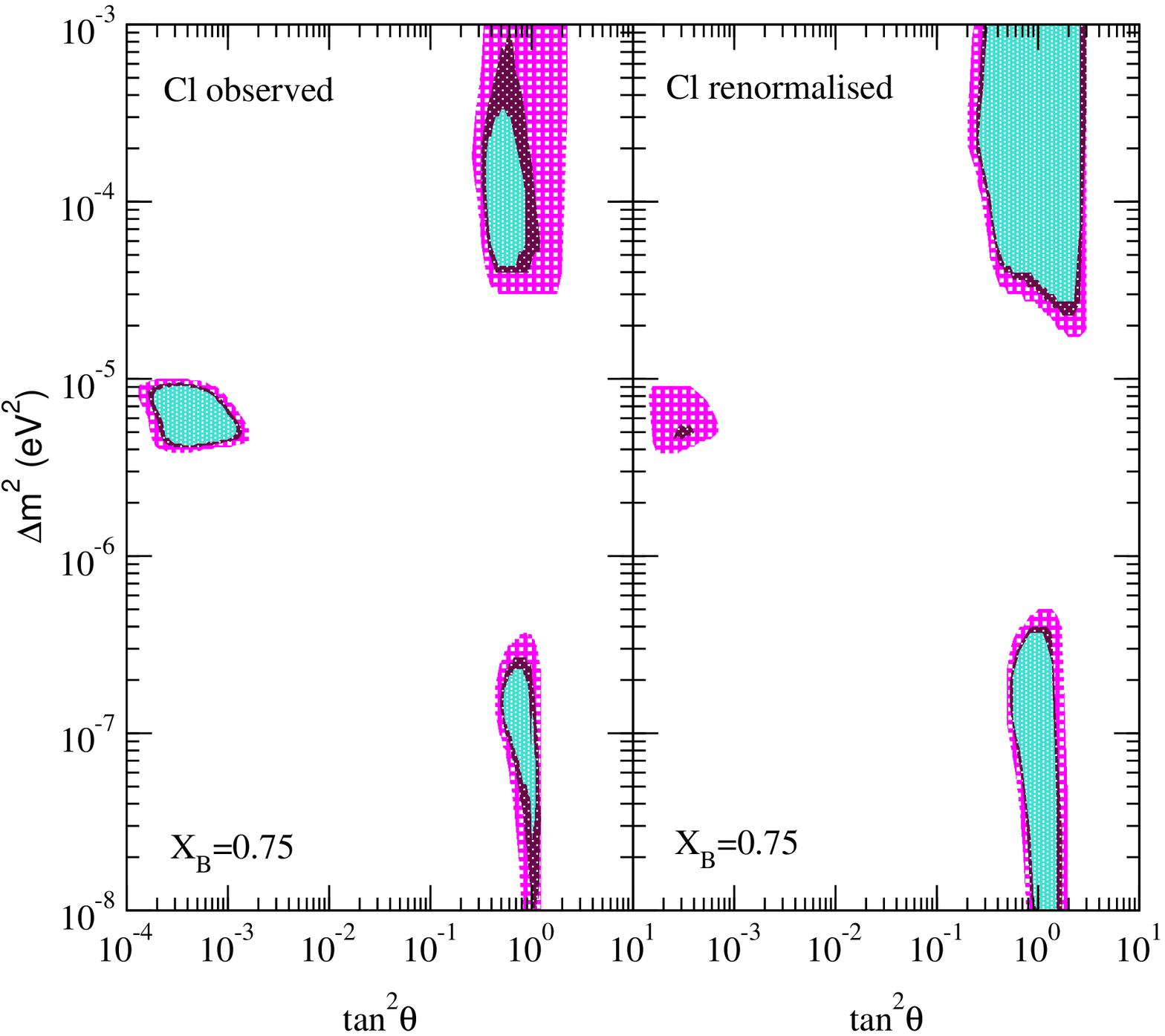,width=16.5cm,height=20.5cm}}
\vskip -1in
\parbox{5in}{
{\bf Fig. 4}: The 90, 95 and 99\% C.L. allowed area from the
global analysis of the total rates from Cl (observed  and
20\% renormalised), Ga and SK detectors
and the 1117 days SK recoil electron spectrum at day and night,
assuming MSW conversions to active neutrinos. The B normalisation
is held fixed at 0.75 of SSM value.}

\end{document}